\documentclass[10pt,twocolumn,letterpaper]{article} 

\usepackage{ol2}
\usepackage[draft]{hyperref}
\usepackage{amsmath}
\usepackage{amssymb}
\usepackage{color}
\usepackage[normalem]{ulem}

\def\el{Electron.\ Lett.\ }

\def\np{Nat.\ Photonics\ }
\def\aop{Adv.\ Opt.\ Photon.\ }
\def\jstqe{IEEE\ J.\ Sel.\ Top.\ Quantum\ Electron.\ }

\begin{document}

\twocolumn[ 

\title{Generation of parabolic similaritons in tapered silicon photonic wires: comparison of pulse dynamics at telecom and mid-IR wavelengths}


\author{Spyros Lavdas,$^{1,*}$ Jeffrey B. Driscoll,$^2$ Hongyi Jiang,$^1$ Richard R. Grote,$^2$ \\ Richard M. Osgood, Jr.,$^2$ and Nicolae C. Panoiu$^1$}

\address{
$^1$Department of Electronic and Electrical Engineering, University College London, Torrington
Place, London WC1E 7JE, UK \\
$^2$Microelectronics Sciences Laboratories, Columbia University, New York, NY 10027, USA

$^*$Corresponding author: s.lavdas@ucl.ac.uk }

\begin{abstract}
We study the generation of parabolic self-similar optical pulses in tapered Si photonic nanowires
(Si-PhNWs) both at telecom ($\lambda=1.55~\mathrm{\mu m}$) and mid-IR ($\lambda=2.2~\mathrm{\mu
m}$) wavelengths. Our computational study is based on a rigorous theoretical model, which fully
describes the influence of linear and nonlinear optical effects on pulse propagation in Si-PhNWs
with arbitrarily varying width. Numerical simulations demonstrate that, in the normal dispersion
regime, optical pulses evolve naturally into parabolic pulses upon propagating in millimeter-long
tapered Si-PhNWs, with the efficiency of this pulse reshaping process being strongly dependent on
the spectral and pulse parameter regime in which the device operates, as well as the particular
shape of the Si-PhNW.
\end{abstract}


 ] 

\noindent Generation of pulses with specific spectral and temporal characteristics is a key
functionality needed in many applications in ultrafast optics, optical signal processing, and
optical communications. One type of such pulses, which can be used as primary information carriers
in optical communications systems, are pulses that preserve their shape upon propagation. Solitons
are the most ubiquitous example of such a pulse that form in the anomalous group-velocity
dispersion (GVD) regime, whereas their counterpart in the normal GVD region are self-similar
pulses, called similaritons \cite{adk93josab,tn96ol,fkt00prl}. Unlike solitons, which require a
threshold power, no constraints have to be imposed on the pulse energy, initial shape, or optical
phase profile to generate similaritons. Due to their self-similar propagation, similaritons do not
undergo wave breaking and the linear chirp they acquire during their formation makes it easy to
employ dispersive pulse compression techniques to generate nearly transform-limited pulses. These
remarkable properties of similaritons have provided a strong incentive for their study, and
optical similaritons have been demonstrated in active optical fiber systems such as Yb-doped fiber
amplifiers \cite{fkt00prl,kph02josab}, using passive schemes based on dispersion-managed or
tapered silica fibers \cite{hn04ol,kbl06el,lts07ol,fpp07oe}, and high-power fiber amplifiers
\cite{lsc02oe,mpp04ol,lls13ol}.

Driven by the ever growing demand for enhanced integration of complex optoelectronic architectures
that process increasing amounts of data, finding efficient ways to extend the regime of
self-similar pulse propagation to chip-scale photonic devices is becoming more pressing. One
promising approach, based on silicon (Si) fibers with micrometer-sized core dimensions
\cite{hss10oe}, has recently been proposed \cite{ph10ol}. A further degree of device integration
can be achieved by employing Si photonic nanowires (Si-PhNWs) with submicrometer transverse size
fabricated on a silicon-on-insulator material system \cite{lll00apl}. In addition to the enhanced
optical nonlinearity and strong frequency dispersion, which allows for increased device
integration, Si-PhNWs allow for seamless integration with complementary metal-oxide semiconductor
technologies. Importantly, the use of Si-PhNWs can be extended to the mid-infrared (mid-IR)
spectral region ($\lambda\gtrsim2.2~\mathrm{\mu m}$) \cite{s10np}, where Si provides superior
functionality due to low two-photon absorption (TPA) and consequently reduced free-carrier
absorption (FCA). In fact, it has already been shown that nonlinear optical effects such as
modulational instability \cite{pco06ol,klo11oe}, frequency dispersion of the nonlinearity
\cite{plo09ol}, and supercontinuum generation \cite{klo11oe,bkr04oe,yla07ol,hcl07oe}, can be used
to achieve significant pulse reshaping in millimeter-long Si-PhNWs (for a review, see
\cite{opd09aop}).

In this Letter, we use a rigorous theoretical model, which describes the propagation of pulses in
Si-PhNWs, and comprehensive numerical simulations to demonstrate that optical similaritons with
parabolic shape can be generated in millimeter-long, dispersion engineered Si-PhNWs. In order to
gain a better understanding of the underlying physics of similariton generation, we present a
comparative analysis of the pulse dynamics in two spectral domains relevant for technological
applications, namely telecom ($\lambda=1.55~\mathrm{\mu m}$) and mid-IR ($\lambda=2.2~\mathrm{\mu
m}$) spectral regions. Thus, the pulse dynamics are described by the following equation
\cite{plo09ol,cpo06jqe,pmw10jstqe,dog12oe}:
\begin{align}
\label{uzt} i\frac{\displaystyle \partial u}{\displaystyle \partial z}&+\sum\limits_{n\geq 1}
\frac{\displaystyle i^{n}\beta_{n}(z)}{\displaystyle n!}\frac{\displaystyle \partial^{n}
u}{\displaystyle \partial t^{n}}= -\frac{\displaystyle ic\kappa(z)}{\displaystyle 2nv_{g}(z)}\alpha_{\mathrm{FC}}(z)u \nonumber \\
&-\frac{\displaystyle \omega \kappa(z)}{\displaystyle n v_{g}(z)}\delta n_{\mathrm{FC}}(z)u
-\gamma(z)\left[1+i\tau(z)\frac{\partial}{\partial t}\right]\vert u\vert^{2} u,
\end{align}
where $u(z,t)$ is the pulse envelope, $z$ and $t$ are the distance along the Si-PhNW and time,
respectively, $\beta_{n}(z)=d^{n}\beta/d\omega^{n}$ is the $n$th order dispersion coefficient,
$\kappa(z)$ quantifies the overlap between the optical mode and the active area of the waveguide,
$v_{g}(z)$ is the group-velocity, $\delta n_{\mathrm{FC}}(z)$ [$\alpha_{\mathrm{FC}}(z)$] are the
free-carrier (FC) induced index change (losses) and are given by $\delta n_{\mathrm{FC}}(z) =
-e^{2}/2\epsilon_{0}n\omega^{2}\left [N(z)/m_{ce}^{*} + N(z)^{0.8}/m_{ch}^{*} \right ]$ and
$\alpha_{\mathrm{FC}}(z) = e^{3}N(z)/\epsilon_{0}cn\omega^{2} (1/\mu_{e}{m_{ce}^{*}}^{2} +
1/\mu_{h}{m_{ch}^{*}}^{2})$, respectively, where $N$ is the FC density, $m_{ce}^{*}=0.26m_{0}$
($m_{ch}^{*}=0.39m_{0}$) is the effective mass of the electrons (holes), with $m_{0}$ the mass of
the electron, and $\mu_{e}$ ($\mu_{h}$) the electron (hole) mobility. The nonlinear properties of
the waveguide are described by the nonlinear coefficient, $\gamma(z)=3\omega
P_{0}\Gamma(z)/4\epsilon_{0}A(z)v_{g}^{2}(z)$, and the shock time scale, \textit{i.e.} the
characteristic response time of the nonlinearity, $\tau(z)=\partial\ln \gamma(z)/\partial \omega$,
where $P_{0}$ is the peak power of the input pulse, and $A(z)$ and $\Gamma(z)$ are the
cross-sectional area and the effective third-order susceptibility of the waveguide, respectively.
Our model is completed by a rate equation describing the FC dynamics,
\begin{align}
\label{re}\frac{\displaystyle \partial N}{\displaystyle \partial t} = - \frac{\displaystyle
N}{\displaystyle t_{c}} + \frac{\displaystyle 3 P_{0}^{2}\Gamma^{\prime\prime}(z)}{\displaystyle
4\epsilon_{0}\hbar A^{2}(z)v_{g}^{2}(z)} |u|^{4},
\end{align}
where $\Gamma^{\prime\prime}$ ($\Gamma^{\prime}$) is the imaginary (real) part of $\Gamma$.
\begin{figure}[b]
\centerline{\includegraphics[width=7.75cm]{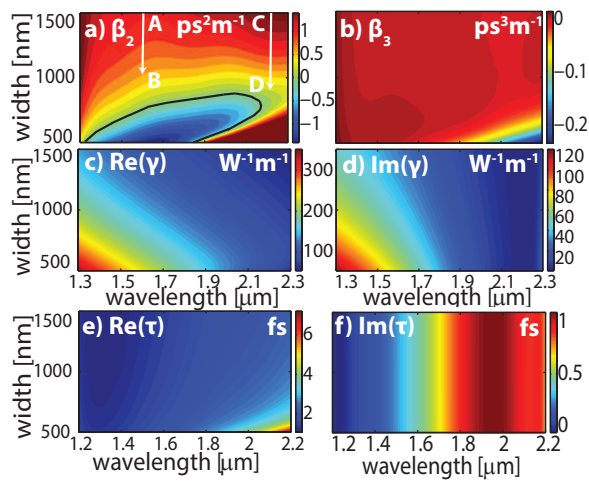}} \caption{Dispersion maps of a)
$\beta_{2}$, b) $\beta_{3}$, c) self-phase modulation coefficient, $\gamma^{\prime}$, d) TPA
coefficient, $\gamma^{\prime\prime}$, and e) real and f) imaginary part of the shock-time
coefficient, $\tau$. In a), $\beta_{2}=0$ on the black contour and arrows indicate the limits of
$w$, at $\lambda=1.55~\mathrm{\mu m}$ and $\lambda=2.2~\mathrm{\mu m}$.} \label{disp}
\end{figure}

The system (\ref{uzt})-(\ref{re}) provides a rigorous description of pulse propagation in Si-PhNWs
with adiabatically varying transverse size since the $z$-dependence of the waveguide parameters is
fully incorporated in our model \textit{via} the implicit dependence of the modes of the Si-PhNW
on its transverse size. Thus, we consider a tapered ridge waveguide with a Si rectangular core
buried in $\mathrm{SiO_{2}}$, with height, $h=250~\mathrm{nm}$, and width, $w$, varying from
$w_{\mathrm{in}}$ to $w_{\mathrm{out}}$ between the input and output facets, respectively. Using a
finite-element mode solver we determine the propagation constant, $\beta(\lambda)$, and the
fundamental TE-like mode, for $1.3~\mathrm{\mu m} \leq \lambda \leq 2.3~\mathrm{\mu m}$ and for 51
values of the waveguide width ranging from $500~\mathrm{nm}$ to $1500~\mathrm{nm}$. The dispersion
coefficients are calculated by fitting $\beta(\lambda)$ with a 12th order polynomial and
subsequently calculating the corresponding derivatives with respect to $\omega$. Using these
results and the corresponding optical modes, the waveguide parameters, $\kappa$, $\gamma$, and
$\tau$, are computed for all values of $w$. The $z$-dependence of these parameters is then
determined by polynomial interpolation.

\begin{figure}[t]
\centerline{\includegraphics[width=7.75cm]{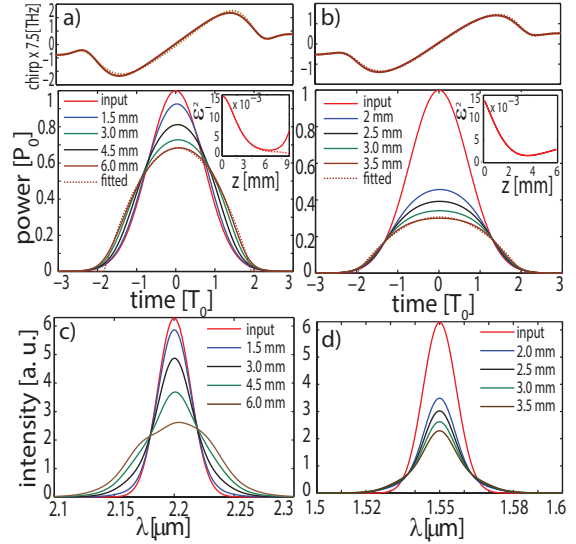}} \caption{Temporal pulse shape with
increasing $z$ and the chirp of the output pulse, calculated for the full model (solid line) and
for $\beta_{3}=0$ and $\tau=0$ (dotted line) (top panels) and the corresponding pulse spectra
(bottom panels). In insets, $\epsilon_I^2$ \textit{vs}. $z$, for the full model (solid line) and
for $\beta_{3}=0$ and $\tau=0$ (dotted line). Left (right) panels correspond to
$\lambda=2.2~\mathrm{\mu m}$ ($\lambda=1.55~\mathrm{\mu m}$).}\label{plsdyn}
\end{figure}
The results of this analysis are summarized in Fig. \ref{disp}, where we plot the dispersion maps
of the waveguide parameters. Thus, Fig. \ref{disp}(a) shows that if $w<887~\mathrm{nm}$ the
Si-PhNW has two zero GVD wavelengths, defined by $\beta_{2}(\lambda,w)=0$, whereas if
$w>887~\mathrm{nm}$ the Si-PhNW has normal GVD in the entire spectral domain. In addition, if
$\lambda>2187~\mathrm{nm}$ the waveguide has normal GVD for any $w$. Important properties of the
Si-PhNW are revealed by the dispersion maps of the nonlinear coefficients as well. Specifically,
the strength of the nonlinearity, $\gamma^{\prime}(\lambda,w)$, decreases with both increasing $w$
and $\lambda$, meaning that in the range of wavelengths and waveguide widths explored here,
nonlinear effects in Si-PhNWs are stronger if narrow waveguides are used at lower wavelengths. On
the other hand the TPA coefficient, $\gamma^{\prime\prime}(\lambda,w)$, and consequently nonlinear
losses, decrease with $w$ and $\lambda$, which suggests that the waveguide parameters and
wavelength must be properly chosen for optimum device operation. Finally, as seen in Fig.
\ref{disp}(e), the shock time $\tau^{\prime}(\lambda,w)$ has large values at long wavelengths but
decreases with $w$.

To investigate the formation of self-similar pulses, we considered first a Gaussian pulse,
$u(t)=e^{-t^{2}/2T_{0}^{2}}$ with full-width at half-maximum (FWHM)
$T_{\mathrm{FWHM}}=220~\mathrm{fs}$ ($T_{\mathrm{FWHM}}=1.665T_{0}$), and peak power
$P_{0}=7~\mathrm{W}$, which is launched in an exponentially tapered Si-PhNW,
$w(z)=w_{\mathrm{in}}e^{-az}$, with $w_{\mathrm{in}}=1500~\mathrm{nm}$. The remaining parameters
are: \textit{i}) at $\lambda=1.55~\mathrm{\mu m}$, $w_{\mathrm{out}}=1080~\mathrm{nm}$,
$\beta_{2,\mathrm{in}}=1.11~\mathrm{ps^{2}m^{-1}}$,
$\beta_{2,\mathrm{out}}=0.79~\mathrm{ps^{2}m^{-1}}$, and length, $L=3.5~\mathrm{mm}$ [arrow $A-B$
in Fig. \ref{disp}(a)], and \textit{ii}) at $\lambda=2.2~\mathrm{\mu m}$,
$w_{\mathrm{out}}=850~\mathrm{nm}$, $\beta_{2,\mathrm{in}}=1.53~\mathrm{ps^{2}m^{-1}}$,
$\beta_{2,\mathrm{out}}=0.088~\mathrm{ps^{2}m^{-1}}$, and $L=6~\mathrm{mm}$ [arrow $C-D$ in Fig.
\ref{disp}(a)]. In Fig. \ref{plsdyn} we plot the pulse profile and its spectrum, calculated for
several values of $z$. As expected, the pulse decay is stronger at $\lambda=1.55~\mathrm{\mu m}$
as compared to that at $\lambda=2.2~\mathrm{\mu m}$, due to larger TPA. The stronger nonlinear
effects at $\lambda=2.2~\mathrm{\mu m}$ are also revealed by the spectral ripples that start to
form at $z\gtrsim5~\mathrm{mm}$ (no such modulations are seen at $\lambda=1.55~\mathrm{\mu m}$).
Also, the pulse becomes more asymmetric at $\lambda=2.2~\mathrm{\mu m}$, due to increased $\tau$
\cite{plo09ol}. However, the most important phenomenon revealed by Fig. \ref{plsdyn} is that at
both wavelengths the pulse evolves into a parabolic one, $\vert u_{p}(t)\vert^{2}=\vert
u_{p}(t_{0})\vert^{2}[1-(t-t_{0})^{2}/T_{p}^{2}]$ for $\vert t-t_{0}\vert<T_{p}$ and $u_{p}(t)=0$
otherwise, where $u_{p}(t_{0})$, $t_{0}$, and $T_{p}$ are the amplitude, time shift, and pulse
width, respectively.

The generation of parabolic pulses can be quantitatively characterized by the intensity misfit
parameter, $\varepsilon_{I}$, which provides a global measure of how close the pulse profile is to
a parabolic one; it is defined as:
\begin{align}
\label{misfit}\varepsilon_{I}^{2} =
\frac{\int[|u(t)|^{2}-|u_{p}(t)|^{2}]^{2}dt}{\int|u(t)|^{4}dt}.
\end{align}
The inset plots in Figs. \ref{plsdyn}(a) and \ref{plsdyn}(b) show that at both wavelengths there
is a certain optimum waveguide length at which $\varepsilon_{I}^{2}$ reaches a minimum value,
namely $\varepsilon_{I}^{2}=1.67\times10^{-3}$ ($\varepsilon_{I}^{2}=1.57\times10^{-3}$) at
$\lambda=2.2~\mathrm{\mu m}$ ($\lambda=1.55~\mathrm{\mu m}$). The small values of
$\varepsilon_{I}^{2}$ provide clear evidence of the formation of parabolic pulses. The pulse
becomes closer to a parabolic pulse at $\lambda=1.55~\mathrm{\mu m}$ because the effects that
induce pulse asymmetry, namely the third-order dispersion and nonlinearity dispersion, are smaller
at this wavelength. This can also be seen by comparing the dependence $\varepsilon_{I}^{2}(z)$ in
the case of the full model (\ref{uzt})-(\ref{re}) and when higher-order effects are neglected
($\beta_{3}=0$ and $\tau=0$). Thus, at $\lambda=1.55~\mathrm{\mu m}$, $\varepsilon_{I}^{2}(z)$ is
almost unaffected if one neglects higher-order effects, whereas in the same conditions, at
$\lambda=2.2~\mathrm{\mu m}$, the minimum of $\varepsilon_{I}^{2}(z)$ decreases considerably to
$0.41\times10^{-3}$ (and is reached at $z=9.52~\mathrm{mm}$).

A fundamental characteristic of parabolic pulses is that across the pulse the frequency chirp
varies linearly with time. The pulses generated in our numerical experiments clearly have this
property, as illustrated in the top panels of Figs. \ref{plsdyn}(a) and \ref{plsdyn}(b). These
figures also show that, at both wavelengths, this linear time dependence of the chirp is preserved
even in the presence of higher-order effects, which demonstrates the robustness against
perturbations of the parabolic pulse generation.
\begin{figure}[t]
\centerline{\includegraphics[width=7.75cm]{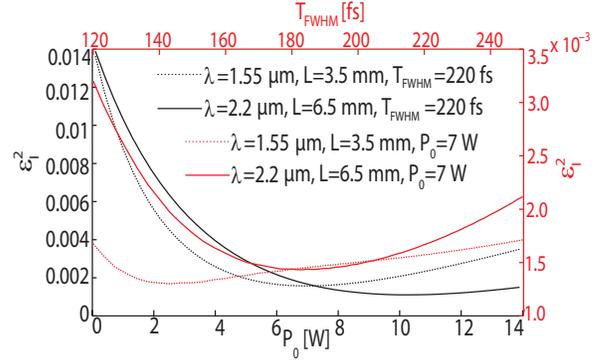}} \caption{Dependence of
$\varepsilon_{I}^{2}$ on pulse width and power.}\label{paramdep}
\end{figure}

The dependence of the similariton generation on the pulse parameters is particularly important
when assessing the effectiveness of this optical process. In order to study this dependence, we
have determined $\varepsilon_{I}^{2}$, at both wavelengths, as a function of pulse parameters,
$T_{\mathrm{FWHM}}$ and $P_{0}$. The results of our analysis, summarized in Fig. \ref{paramdep},
show that for a given waveguide length there is an optimum power at which $\varepsilon_{I}^{2}$
reaches a minimum, which is explained by the fact that the similariton formation length increases
with $P_{0}$. By contrast, there is no optimum value of $T_{\mathrm{FWHM}}$ at which
$\varepsilon_{I}^{2}$ becomes minimum.

\begin{figure}[b]
\centerline{\includegraphics[width=7.75cm]{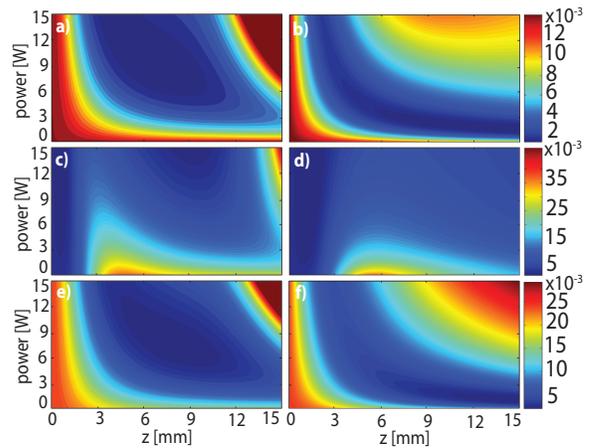}} \caption{Dependence of
$\varepsilon_{I}^{2}$ on $z$ and pulse power, calculated for a Gaussian pulse (top panels),
supergaussian with $m=2$ (middle panels), and sech pulse (bottom panels). In all cases
$T_{\mathrm{FWHM}}=220~\mathrm{fs}$. Left (right) panels correspond to $\lambda=2.2~\mathrm{\mu
m}$ ($\lambda=1.55~\mathrm{\mu m}$).}\label{profdep}
\end{figure}
The relation between the input pulse parameters and the similariton generation can be further
explored by considering pulses with different shapes. Our results regarding this dependence are
summarized in Fig. \ref{profdep}, where we plot the evolution of $\varepsilon_{I}^{2}(z)$,
determined for varying $P_{0}$. As input pulses we considered a Gaussian pulse, a supergaussian
pulse, $u(t)=e^{-t^{2m}/2T_{0}^{2m}}$ with $m=2$ ($T_{\mathrm{FWHM}}=1.824T_{0}$), and a sech
pulse, $u(t)=\mathrm{sech}(t/T_{0})$, where $T_{\mathrm{FWHM}}=1.763T_{0}$. In all cases
$T_{\mathrm{FWHM}}=220~\mathrm{fs}$. There are several revealing conclusions that can be drawn
from the maps in Fig. \ref{profdep}. First, the Gaussian pulse leads to the lowest values of
$\varepsilon_{I}^{2}(z)$, which suggests that this pulse shape is the most efficient one for
generating similaritons. Second, in the case of Gaussian and sech pulses there is a band of low
values of $\varepsilon_{I}^{2}(z)$, which is narrower at $\lambda=1.55~\mathrm{\mu m}$ as compared
to its width at $\lambda=2.2~\mathrm{\mu m}$ and in both cases it broadens as $P_{0}$ decreases,
whereas in the case of supergaussian pulses two such bands exist. Finally, pulses with a
supergaussian shape evolve into a similariton over the shortest distance, which is explained by
the fact that of the three pulse profiles the supergaussian one is closest to a parabolic pulse.

\begin{figure}[t]
\centerline{\includegraphics[width=7.75cm]{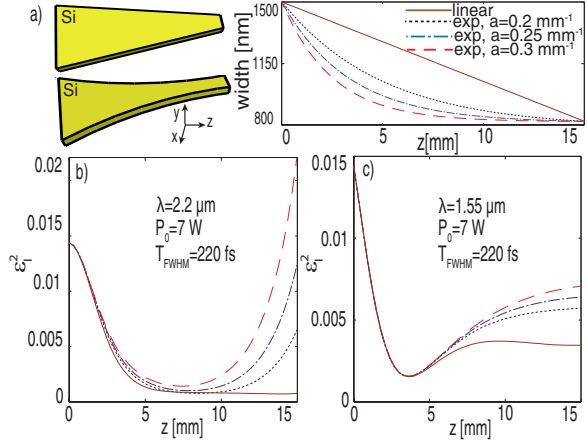}} \caption{a) Schematics and dependence
$w(z)$ for a linear taper and exponential ones,
$w(z)=w_{\mathrm{in}}+(w_{\mathrm{out}}-w_{\mathrm{in}})(1-e^{-az})/(1-e^{-aL})$. In all cases
$w_{\mathrm{in}}=1500~\mathrm{nm}$ and $w_{\mathrm{out}}=820~\mathrm{nm}$. b) and c) show the
evolution of $\varepsilon_{I}^{2}$ \textit{vs}. $z$, for the tapers in a), at
$\lambda=2.2~\mathrm{\mu m}$ and $\lambda=1.55~\mathrm{\mu m}$, respectively.}\label{taperdep}
\end{figure}
Due to its practical relevance, we also studied the generation of similaritons in Si-PhNW tapers
with different profiles. To this end, we considered a linear taper and exponential ones with
different $z$-variation rate, in all cases the (Gaussian) pulse parameters and $w_{\mathrm{in}}$
and $w_{\mathrm{out}}$ being the same [see Fig. \ref{taperdep}(a)]. The results of this analysis,
which are presented in Fig. \ref{taperdep}, show that although similaritons are generated
irrespective of the taper profile, the efficiency of this process does depend on the shape of the
taper. In particular, overall the linear taper is the most effective for similariton generation,
whereas in the case of exponential tapers the steeper their profile the more inefficient they are.
These conclusions qualitatively remain valid at both $\lambda=2.2~\mathrm{\mu m}$ and
$\lambda=1.55~\mathrm{\mu m}$, although the overall pulse dynamics do depend on wavelength. In
particular, $\varepsilon_{I}^{2}$ is smaller at $\lambda=2.2~\mathrm{\mu m}$ and the pulse
preserves a parabolic shape for a longer distance, in agreement with the results in Fig.
\ref{profdep}.

In conclusion, we have demonstrated that parabolic pulses can be generated in millimeter-long
tapered Si-PhNWs with engineered decreasing normal GVD. Our analysis showed that using this
approach optical similaritons can be generated at both telecom and mid-IR wavelengths,
irrespective of the pulse shape and taper profile. However, our investigations have revealed that
the efficiency of the similariton generation is strongly dependent on the wavelength at which the
device operates, pulse parameters and its temporal profile, as well as the particular shape of the
Si-PhNW taper.

The work of S.L. was supported through a UCL Impact Award graduate studentship. R. R. G.
acknowledges support from the Columbia Optics and Quantum Electronics IGERT under NSF Grant
DGE-1069420.

\end{document}